\documentclass[]{aa}  

\usepackage{graphicx}
\usepackage{txfonts}
\usepackage{lipsum}
\usepackage{subcaption}         
\usepackage{lscape}             
\usepackage{placeins}           

\usepackage{amsmath,amssymb}
\begin{document}

\nolinenumbers  
%
%
%
\title{Convective Heat Transport Asymptotic Law in an $\alpha^2$ dynamo model}
\author{G. Nigro\inst{1}}
\institute{Dipartimento di Fisica, Universit\`a degli Studi di Roma Tor Vergata, Via della Ricerca Scientifica 1, Roma, 00133, Italy \\
\email{giuseppina.nigro@roma2.infn.it, giusy.nigro@gmail.com} }
%
\date{today}
%
\abstract
{Stellar activity and planetary magnetospheres are powered by an underlying dynamo mechanism generated by magnetoconvection coupled with rotation. In astrophysical contexts, magnetoconvection typically occurs in parameter regimes that are currently inaccessible to direct numerical simulations (DNS).}
%
{We investigate convective heat transfer in a magneto-convection and dynamo model under extreme parameter conditions, specifically high Rayleigh and Prandtl numbers, in a plasma flow with maximum kinetic helicity compatible with fast rotating objects.}
%
{Our approach to studying magneto-convection and dynamo mechanisms employs a simplified thermally driven shell model. Magnetic polarity reversals are obtained by including a pitchfork bifurcation term in the large-scale magnetic field equation, while nonlinear dynamics are described by a shell model formulation. The low computational cost of the model allows us to explore the asymptotic behavior of convective heat transfer in regimes beyond those reached by current DNS.}
{Our results reveal that the Nusselt number $Nu$ -- a dimensionless measure of convective heat transport-- generally increases with turbulence, following a power-law scaling and showing a strong correlation with Ra and Pr. This relationship appears to be more pronounced than that observed in non-magnetized fluids, suggesting that magnetic fields may significantly enhance convective heat transfer.}
{Despite the assumption to neglect spatial information such as density stratification -- an assumption that is necessary in the shell model approach -- our model captures the gross dynamical features of turbulent magnetoconvection in asymptotic regimes. It allows for a broad exploration of parameter space, indicating that magnetic fields may play a central role in modulating heat transport in stellar and planetary interiors.}


\maketitle


\section{Introduction}
\label{Intro}
Stars and planets are typically characterized by magnetic fields generated through the magnetic dynamo, namely a mechanism that converts kinetic energy into magnetic energy via magnetoconvection in an electrically conductive fluid, such as a stellar plasma, powered by rotation. Basically, helical motion, produced by the combined effects of convection and rotation, ensures the breaking of mirror symmetry that guarantees the amplification and maintenance of significant magnetic fields \citep{Moffatt_1978}. 
Although the specific configuration of the dynamo mechanism depends on the internal structure of the convective zone, general features that are common across various stellar types and some planets can be identified \citep{Christensen_et_al_2009Natur}.

In stars with a radiative core, differential rotation often develops within the convection zone, along with a velocity shear layer at the boundary between the radiative core and the convection envelope, similar to the solar tachocline. In contrast, fully convective stars lack a radiative zone and usually display nearly solid-body rotation \citep[e.g.][]{Donati_et_al_Sci_2006, Morin_et_al_2008MNRAS_a, Davenport_et_al_AJ_2020}. This evidence provides the possibility to distinguish different scenarios for the dynamo in stars that develop differential rotation and stars that show differential rotation compatible with solid-body rotations. Specifically, when differential rotation is significant, it acts to produce the $\Omega$ effect that, together with the helical turbulence ($\alpha$ effect), generates a kind of dynamo classified as the $\alpha$-$\Omega$ dynamo. In contrast, when differential rotation does not play a significant role, the mechanism is classified as the $\alpha^2$ dynamo. 
In an $\alpha^2$ dynamo, the magnetic field is amplified primarily through cyclonic turbulence generated by magnetoconvection in conjunction with stellar rotation. The latter type of dynamo is conjectured to be responsible for the mechanism of magnetic field generation in fully convective stars and planets like the Earth, as observations are compatible with a lack of velocity shear and differential rotation in their convection zones.

In a simplified framework, an $\alpha^2$ dynamo can be modeled through Rayleigh-B {e}rnard (RB) convection in a magnetohydrodynamic (MHD) fluid that, under the effect of rotation, produces helical flow capable of amplifying and sustaining intense magnetic fields. RB convection is a paradigm of a fluid layer heated from the bottom and cooled at the top, where the Nusselt number $Nu$ is a dimensionless parameter that measures the transfer of convective heat within the fluid from the bottom to the top boundary layer \citep{Ozisik_1985, White_1988}.

The challenge of characterizing heat transport in astrophysical convection arises from the enormous range of scales involved. In the solar convection zone, merely as an illustrative example, spatial scales span from giant cells $\sim200$ Mm near the base (lifetimes $\sim1$ month) to granulation $\sim1$ Mm and sub-granular features down to tens of kilometers at the solar surface evolving in minutes \citep{Hathaway_et_al_Science_2013, Miesch_LRSP_2005, Nordlund_Asplund_LRSP_2009, Kuridze_et_al_apJ_2025}. 
Consequently, the enormous range of spatial and temporal scales implies an extremely high Reynolds number, possibly reaching $Re \sim 10^{15}$ or even higher. Specifically, \citet{Rieutord_Rincon_2010LRS} consider extreme convection parameters, such as the range of Re from $10^{10}$ to $10^{13}$ in the solar convection zone, based on the (large) vertical extent of the convective layer and typical convective velocities. \citet{Hanasoge_et_al_2012PNAS} mention parameter values such as Re $\sim 10^{12}-10^{16}$ and Ra $\sim 10^{19}-10^{24}$. We expect similar values for the convection zone in other late-type stars. 

Unfortunately, the convection regimes observed in astrophysical contexts are not yet accessible to current direct numerical simulations. To probe the asymptotic regime of magnetoconvection, where both turbulent and magnetic Reynolds numbers are very large, one must therefore rely on reduced or parameterized models. Simplified dynamical models—such as mean-field, low-order, or shell-model formulations—offer an alternative route to explore convection and dynamo processes under extreme parameter conditions. Although shell models neglect geometric and stratification effects \citep{Biferale_2003}, they retain key nonlinear interactions and allow for a wide exploration of parameter space at relatively low computational cost. In particular, shell models, representing turbulent cascades as a hierarchy of interacting Fourier shells, are well suited to studying asymptotic trends and scaling laws \citep[e.g.][]{Frick_Sokoloff_1998, Biferale_2003, Nigro_2022}.

%
Understanding how $Nu$ asymptotically scales on the control parameters of magnetoconvection, such as the Rayleigh (Ra) and Prandtl (Pr) numbers, is crucial because this scaling may provide a quantitative measure of the efficiency of convective heat transport in the regimes which are pertinent to naturally occurring dynamos. In stellar interiors, where convection is highly supercritical, the resulting turbulent flows might yield large {\it Nu}, indicating efficient heat transport and reflecting the dominance of convective over radiative and conductive transport \citep{Brun_Browning_2017}. 
In hydrodynamic convection, $Nu$ follows power-law relationships with $Ra$ and the Prandtl number ($Pr$), such as $Nu \propto Ra^{\beta} Pr^{\gamma}$, with exponents that depend on the turbulence regime \citep{Ahlers_et_al_2009, Chilla_et_al_2012, Pandey_Sreenivasan_EPL_2021}. However, the corresponding scaling laws in magnetized convection remain largely unexplored. 
In magnetohydrodynamic simulations of mangnetoconvection \citep[][e.g.]{Brun_et_al_ApJ_2022}, part of the star’s energy (luminosity) is converted into convective energy flux and magnetic energy by nonlinear dynamos, suggesting that the Nusselt number may also capture magnetically mediated energy transport. 
Therefore, the Nusselt number may not only quantify the efficiency of convective heat transport but also reflect the fraction of energy redistributed by magnetic processes. These aspects are crucial for understanding how convective heat transfer can be coupled with magnetic field amplification and the emergence of large-scale dynamo action. In line with this view, global 3D simulations \citep[e.g.][]{Miesch_et_al_ApJ_2000, Brun_et_al_ApJ_2004, Christensen_Aubert_2006, Browning_et_al_ApJ_2006, Christensen_et_al_2009Natur, Brun_Browning_2017} have emphasized the central role of convection in regulating magnetic field generation through dynamo processes.

In fully convective stars, because of the absence of a radiative zone, the entire stellar luminosity must be carried by convection; hence, $Nu$ becomes a key diagnostic of the vigor of convective flows capable of sustaining stellar luminosity \citep{Brun_et_al_ApJ_2004, Browning_2008ApJ, Brun_et_al_ApJ_2022} and magnetic fields \citep{Christensen_Aubert_2006, Christensen_et_al_2009Natur, Nigro_2022}. 

Previous studies in nonmagnetized RB convection have established that $Nu$ generally follows power-law relations with $Ra$ and the Prandtl number ($Pr$), $Nu \propto Ra^\beta Pr^\gamma$, where the exponents $\beta$ and $\gamma$ depend on the flow regime \citep{Ahlers_et_al_2009, Chilla_et_al_2012, Pandey_Sreenivasan_EPL_2021}. However, the extension of these scaling laws to magnetized flows remains largely unexplored. 
Our approach offers insights into how convective heat transport can behave in an MHD fluid at extreme parameter values and how it is coupled with magnetic field amplification when density stratification, $\Omega$ effect and other inhomogeneity and spatial properties are neglected. 
Our approach relies on a thermally driven shell model to describe magnetoconvective dynamics and the $\alpha^2$ dynamo in the asymptotic regime \citep{Nigro_2022}. 

The article is organized as follows: Section \ref{numerical_model} describes the model, Section \ref{results} presents the numerical results and the derived scaling laws, and Section \ref{conclusions}, summarizes our findings and discusses them in the context of previous studies on the asymptotic law of convective heat transport in hydrodynamic fluids.

\section{Numerical Model}
\label{numerical_model}

The model adopted in this study is a thermally driven magnetoconvective shell model, introduced by \citet{Nigro_2022}, which is described by the following coupled equations: 

\begin{align}
&\left(\frac{d}{dt} + \nu k_n^2\right) u_n = - \tilde{\alpha} \theta_n \;
+ {i} k_n\;[(u_{n+1}u_{n+2}-b_{n+1}b_{n+2}) + &
\nonumber
\\
&- \frac{\xi}{2}(u_{n-1}u_{n+1}-b_{n-1}b_{n+1})
- \frac{1-\xi}{4}(u_{n-2}u_{n-1}-b_{n-2}b_{n-1})]^* 
\label{eq:GOY_u}
\\
&\left(\frac{d}{dt} + \eta k_n^2\right) b_n = 
{i} k_n \; [(1-\xi - \xi_m)(u_{n+1}b_{n+2}-b_{n+1}u_{n+2})+
\nonumber
\\
&+ \frac{\xi_m}{2}(u_{n-1}b_{n+1}-b_{n-1}u_{n+1}) + 
\frac{1-\xi_m}{4}(u_{n-2}b_{n-1}-b_{n-2}u_{n-1})]^*
\label{eq:GOY_b}
\\
&\left(\frac{d}{dt} + \chi k_n^2\right) \theta_n =
{i} k_n [\alpha_1 u_{n+1}^*\theta_{n+2}^* + \alpha_2 u_{n+2}^* \theta_{n+1}^* +
\nonumber
\\
& + \beta_1 u_{n-1} \theta_{n+1} - \beta_2 u_{n+1} \theta_{n-1} 
+ \gamma_1 u_{n-1} \theta_{n-2} + \gamma_2 u_{n-2} \theta_{n-1}]^* + f_n
\label{eq:Thermal}
\end{align}

where $u_n$ are the velocity field fluctuations, $b_n$ magnetic field fluctuations, and $\theta_n$  temperature fluctuations at the shell $n$, where $n=1, \dots, N$ and $N$ is the total number of shells, $\nu$ is the kinematic viscosity, $\eta$ the magnetic diffusivity, and $\chi$ is the thermal diffusivity, finally $\tilde{\alpha}$ denote the thermal convection coefficient. The values of the model coefficients $\xi = 1/2$ and $\xi_m = 1/3$ are chosen in such a way that equation (\ref{eq:GOY_u}) and (\ref{eq:GOY_b}) conserve the quadratic invariants of the MHD equations -- total energy, cross helicity, and magnetic helicity -- in the ideal case, namely where there is no external forcing or dissipation, and when $\tilde{\alpha} = 0$ \citep{Frick_Sokoloff_1998, Biferale_2003}. The values of the coefficients in the temperature equations are $\alpha_1 = \alpha_2 = 1$, $\beta_1 = \beta_2 = 1/2$, and $\gamma_1 = \gamma_2 = - 1/4$ such that we adopt the coupling with the model of \citet{Jensen_et_al_1992} (see also \citet{Mingshun_Shida_1997}). Finally, the forcing $f_n$ is a random Gaussian signal with unit standard deviation, self-correlated with a correlation time of 1 unit time. The equations are normalized to the typical physical parameters, namely the velocity is normalized to the freefall velocity $U = \sqrt{\alpha g L \Delta T} = \sqrt{\nu \chi \textrm{Ra} }/L $, considering the typical magnetic field $ B_0 = U$ (the typical magnetic field obtained when the freefall velocity $U$ is equal to the Alfv\'en speed), the temperature is measured in terms of $\Delta T$; and finally, the length in units of $L$, that is, the characteristic length of the system and the time in units of freefall time $( L/ \hat{\alpha} g \Delta T)^{1/2}$. 

The shell model describes the evolution of velocity, temperature and magnetic field fluctuations in a hierarchy of exponentially spaced wavenumber shells $k_n = k_0 \lambda^n$, with $n = 1, \dots, N$, where $\lambda=2$ is the shell-spacing parameter. Each shell represents a scale of turbulent eddies, and the nonlinear interactions are constructed to mimic the local cascade processes in magnetohydrodynamic (MHD) turbulence. 

The large-scale magnetic field ${b}_1$ is modified to include a supercritical pitchfork bifurcation term \citep{Nigro_Carbone_2011}:
\begin{equation}
\frac{d b_1}{dt}= - \eta k_1^2 b_1 +
{ i} \frac{k_1}{6}\; (u_{2}^*b_{3}-b_{2}^*u_{3})+ \mu b_1 \left( 1- \frac{b_1^2}{B_0^2}\right) \ .
\label{eq:b1}
\end{equation}
The pitchfork bifurcation term accounts for magnetic field reversals, as $b_1 = \pm B_0$ are two equilibrium states for the large-scale magnetic field $b_1$, while $b_1=0$ is an unstable equilibrium (for more details, see \citet{Nigro_2022}). 

The system is numerically evolved using the fourth-order Runge-Kutta numerical scheme, and considering a wide range of model parameters, particularly focusing on varying the thermal diffusivity $\chi$, kinematic viscosity $\nu$ and magnetic diffusivity $\eta$, which control the values of the Rayleigh (Ra$= {\tilde{\alpha}}{\theta_0}L^3/(\nu \chi)$, with $\theta_0 =  \sqrt \langle {\sum_{n=1}^N \theta_n^2}\rangle_{t}$), Reynolds (Re$=Lu_0/\nu$, where $u_0 = \sqrt \langle {\sum_{n=1}^N u_n^2} \rangle_{t}$) and magnetic Reynolds (Rm$=Lu_0/\eta$) numbers, respectively. The Prandtl number (Pr) and the magnetic Prandtl number (Pm) are likewise determined by the ratios $\nu/\chi$ and $\nu/\eta$.

The objective of this model is to reproduce the essential features of a $\alpha^2$ dynamo in a convective MHD plasma with a large kinetic helicity that we assume is produced by fast rotation, i.e., the ratio between kinetic helicity and kinetic energy is equal to 1 (maximum kinetic helicity.

An important feature of this model is its computational efficiency, which enables exploration of turbulent regimes with Ra, Re, and Rm values well beyond the reach of current direct numerical simulations, although spatial information is neglected (as density stratification) according to the shell model assumptions. The model captures essential aspects of convective heat transfer and dynamo action, including magnetic-field reversals and their relationship to the level of turbulence and convective transport.

\section{Numerical Simulations}
\label{results}

\begin{figure*}
   \includegraphics[width=18cm]{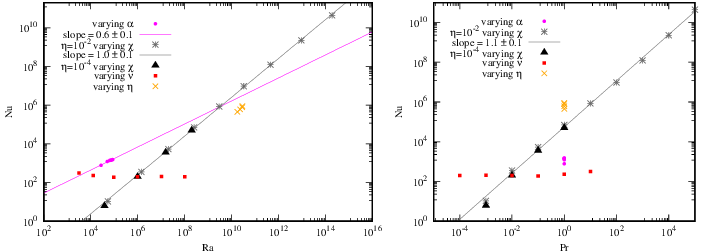}
    \caption{The asymptotic behavior of the Nusselt number $Nu$ as the Rayleigh number Ra and the Prandtl number Pr increase.}
    \label{fig:Nusselt}
\end{figure*}

The Nusselt number $Nu$ is a dimensionless measure of convective heat transport expressed in terms of the ratio of total heat flux (convective and conductive) and conductive heat flux \citep{Ahlers_et_al_2009, Chilla_et_al_2012, Verma_book, Pandey_et_al_2021}. Taking into account $v_z$ as the velocity component along the acceleration of gravity, i.e., along the $z$ axis, $u$ as the vertical turbulent velocity and $\Theta$ the temperature fluctuations, following  \citet{Benzi_et_al_1998}, we can write the following: 
\begin{align}
Nu = \frac{\langle  v_z  T \rangle - \chi \frac{ \partial{\langle T \rangle} }{ \partial z}  }{  \chi \frac{\Delta T}{L} }
= \frac{\langle u \Theta \rangle}{ \chi \frac{\Delta T}{L} } \ ,
\label{eq:Nu1}
\end{align}
where the brackets represent the space and time averages. 

In direct numerical simulation (DNS), there are different approaches to calculate $Nu$, although this should converge to a consistent value if the DNS is well resolved. Crucial, therefore, is the definition adopted to compute $Nu$ as it concerns the characterization of heat transport. For example, \citet{Xu_et_al_2020} tested different approaches to computing $Nu$, showing that the relation $Nu = \sqrt{Ra Pr} \langle u \Theta \rangle_{V,t}$, where $w$ is the velocity component along ${\bf g}$, and the brackets represent the volume $V$ and the time average, provides the best definitions, among those they tested, to better describe the heat transport in their setup. 
In our shell model framework, the Nusselt number is defined consistently with the model formalism as
\begin{align}
Nu = \sqrt{\textrm{Ra Pr}} \; \left\langle \sum_{n = 1}^N u_n \theta_n \right\rangle_t \ ,
\label{eq:Nu2}
\end{align}
where Pr is the Prandtl number and the brackets $\langle \rangle_t$ indicate a time average, the velocity is normalized to the free fall velocity $U = \sqrt{\alpha g L \Delta T} = \sqrt{\nu \chi Ra}/L $, and the temperature to $\Delta T$.


Although our model provides a limited description of magnetoconvection, it allows us to investigate the asymptotic behavior of $Nu$ in regimes of extreme parameters. 
Figure~\ref{fig:Nusselt} shows the dependence of $Nu$ on the Rayleigh and Prandtl numbers as the model parameters vary individually.
In this model, $Nu$ shows comparable dependencies on $Ra$ and $Pr$, approximately following $Nu \propto$ Ra and $Nu \propto $ Pr, as shown in the left panel and right panel, respectively, of Fig.~\ref{fig:Nusselt} when $\chi$ varies. 
Different asymptotic behaviors emerge when the dissipative coefficients and the thermal expansion parameter $\tilde{\alpha}$ are varied independently.

We focus our investigation mainly on the dependence of $Nu$ on the thermal diffusivity $\chi$, which changes the values of both Ra and Pr, and we treat the dependence on the other parameters as an indication, which requires further investigation to be conclusive. Furthermore, the magenta line in the left panel of Fig.~\ref{fig:Nusselt} is just the best fit of the points limited in our data, and we do not pretend to extrapolate the power-law behavior of this curve to the very large Rayleigh numbers shown in the plot.

We find that $Nu$ exhibits a clear power-law dependence on Ra and Pr, consistent with previous studies. However, the slopes of such power laws are different from those found in the literature. This can be due to many different reasons, and we want to explore them here in more detail. 

To better interpret these results, it is useful to recall the classical studies on the asymptotic laws of convective heat transport in non-magnetic Rayleigh–Bénard convection. 
Since the pioneering works by \citet{Malkus_54a, Malkus_54b} and \citet{Spiegel_1962}, the scaling of the Nusselt number with the Rayleigh number has been the subject of extensive theoretical and experimental investigation, sometimes with contradictory results (see for more discussion \citet{Ahlers_et_al_2009, Chilla_et_al_2012, Verma_book} and references therein). 
Indeed, a great deal of literature has been produced on this issue \citep[e.g.][]{Spiegel_1962a, Spiegel_1971, Threlfall_1975, Castaing_et_al_1989, Kerr_periodicBC_1996, Niemela_et_al_2006, Pandey_2021} since the oldest works made by \citet{Malkus_54a, Malkus_54b, Spiegel_1962}, who derived $Nu \sim {\textrm{Ra}}^{1/3}$ based on perturbation analysis and their laboratory experiment and \citet{Kraichnan_1962}'s scaling law. Kraichnan derived the rigorous upper bound for turbulent heat transport $Nu \sim {\textrm{Ra}}^{1/2}$ by applying the variational calculus to the hydrodynamic Boussinesq equation (see also \citet{Spiegel_1971}). 
More recently, \citet{Niemela_et_al_2000, Niemela_et_al_2006} have found $Nu = 0.088 \;{\textrm{Ra}}^{0.32}$ in a very broad range of the asymptotic regime regarding global heat transport within a cryogenic helium gas. This gas was chosen as a nonmagnetic fluid for an experiment investigating the asymptotic behavior of turbulent fluid convection. Even more recently, \citet{Iyer_et_al_2020} have shown in three-dimensional hydrodynamic turbulent convection that the heat transport continues to follow the classical 1/3 scaling law (see also \citet{He_et_al_2012, Urban_et_al_2012, Doering_et_al_2019}). 

Subsequent high-precision measurements proved that $Nu$ depends not only on Ra, but also on Pr. In particular, $Nu \sim {\textrm{Pr}}^{\gamma}$ where the exponent $\gamma$ takes different values for different Ra numbers \citep{Bhattacharya_et_al_2021}. Recently, \citet{Pandey_Sreenivasan_EPL_2021} have found $Nu \approx 6 \; {\textrm{Ra}}^{1/3} {\textrm{Pr}}^{1/6}$ in direct numerical simulations of an RB convection flow. It is important to note that the weak dependence of $Nu$ on Pr made it challenging to catch the precise dependence during the first years after Malkus-Spiegel's work. This weak dependence, together with the difficulty of reaching asymptotic regimes, sometimes generated controversy over the exact scaling of $Nu$ with Ra \citep{He_et_al_2012, Urban_et_al_2012, Doering_et_al_2019}. Nevertheless, even if the $Nu$ dependency on Pr has turned out to be very small, this can be relevant for solar and stellar magneto-convection, as the Pr values are very small. Nevertheless, it was evident that $Nu$ has shown a dependence not only on Ra and Pr, but also on how turbulent velocity and temperature fluctuations are correlated, which, in turn, depends on the particular state of the turbulence. For this reason, in previous decades, figuring out the exact relationship $Nu = Nu\;({\textrm{Ra, Pr}})$ in hydrodynamic fluids has not been easy. 

These classical results form a reference point for interpreting our findings. However, in our magnetized system, additional effects, such as magnetic feedback and the absence of explicit geometry, shape the asymptotic behavior of $Nu$. Indeed, it is important to bear in mind that our model does not allow any particular geometrical description owing to the shell model technique, whereas some studies, even though not extreme parameter regimes, have shown how geometric constraints could have some influence on $Nu$ asymptotic behavior in non-magnetized RB convection \citep{Niemela_et_al_2006, Wagner_2013, Huang_et_al_2013, Chong_et_al_2015, Chong_Xia_2016}. 
In fact, it has been proven that the particular geometry of the nonmagnetised flow is relevant for heat transfer, where large-scale structures of the convection cell pattern can usually play a dominant role \citep[e.g.][]{Sugiyama_et_al_2010, Chandra_et_al_2011_PhRvE, Chandra_et_al_2011_JPhCS, Chandra_Verma_2013, Xi_et_al_2016, Xu_et_al_2020}. 
Several studies have, in fact, shown a connection between $Nu$ and flow structures
\citep{Sun_et_al_2005, Xi_et_al_2008, Weiss_Ahlers_2011, van_der_Poel_2011}. In particular, \citet{Xi_et_al_2016} in a non-magnetized RB experimental setup of a 3D cylindrical container observed a temporary $Nu$ increase during the flow reversal due to a more coherent flow and plumes, which increases the heat transfer efficiency, contrary to a cessation during which $Nu$ drops to its local minimum. 

Another important remark is that all the studies mentioned above are purely hydrodynamic and do not take into account the effects of the magnetic field on the flow. This can, in our opinion, be a crucial point, as the presence of a magnetic field is known to drastically change the behavior of basically any fluid instability \citep{Chandrasekhar_book}. In fact, to our knowledge, our study is the first to examine the asymptotic behavior of $Nu$ as a function of Ra and Pr in the magnetized case. Indeed, studying the asymptotic behavior of an MHD flow in direct numerical simulations and laboratory experiments remains a very challenging task. However, shell models, despite their important approximations, allow exploration of extreme parameter regimes while retaining many features of an MHD flow.   

Considering the asymptotic regimes obtained for increasing global $Nu$, our model, which includes magnetic field dynamics coupled with convection, shows a clear tendency towards a higher frequency of magnetic-polarity reversals \citep{Nigro_2022}. 
This link between enhanced convective transport and magnetic variability resembles the behavior observed in purely hydrodynamic systems, where bursts in heat transport often precede large-scale circulation reversals. However, unlike hydrodynamic systems, the reversals observed in our model affect the magnetic field polarity, highlighting the central role of convective heat transport in favoring such magnetic transitions.
Specifically, considering the analogy between large-scale circulation reversals observed in thermal convection and magnetic polarity reversals observed in the fluid dynamo, we have shown that a rapid increase in convective heat flux, i.e., instantaneous $Nu$, can cause the onset of a magnetic polarity reversal \citep{Nigro_2022} in the physical situations reproduced by our model.
In addition, simulations that achieve average global $Nu$ values higher than those of the other simulations are characterized by more frequent magnetic reversals and thus significantly higher magnetic variability (a detailed analysis of this aspect is in preparation). 
However, it turns out that the growth trend in the number of magnetic polarity reversals with $Nu$ is complicated by the dependence of $Nu$ on other parameters, such as Rm and Re. In general, increasing the turbulence level leads to a higher frequency of magnetic reversals (see Fig.~3 in \citep{Nigro_2022}).

It is important to note that high levels of turbulence do not necessarily imply enhanced convective heat transport. What ultimately determines the efficiency of heat transfer is the degree of correlation between temperature and velocity fluctuations. In certain turbulent regimes, as shown in previous studies of fluid convection, these two fields may become partially decorrelated, thus reducing the effective heat flux \citep[e.g.,][]{Pandey_Sreenivasan_EPL_2021}.
In particular, simulations showing a relatively constant behavior of $Nu$ as a function of Ra and Pr (simulation represented by red squares in Figure \ref{fig:Nusselt}) are cases in which increasing turbulence levels does not substantially increase the transport of convective heat. 
However, in most of our simulations, the Nusselt number, which is the correlation between the temperature and velocity field fluctuations, scales with the turbulence level as $Nu \propto$ Re, Rm. 

\section{Discussions and conclusions}
\label{conclusions}

We have investigated the asymptotic behavior of convective heat transport in a simplified magneto-convective system based on a thermally driven shell model designed to capture the essential dynamics of an $\alpha^2$ dynamo. The main advantage of this approach lies in its ability to access extreme parameter regimes, characterized by very high values of Ra, Re, and Rm, while maintaining reasonable computational cost. This makes it particularly suited for investigating asymptotic behaviors that are otherwise inaccessible via direct numerical simulations.

The model reproduces several key features of turbulent magnetized convection, such as the emergence of magnetic polarity reversals and their connection with enhanced convective heat transport \citep{Nigro_2022}. The scaling behavior of the Nusselt number $Nu$ with the Rayleigh and Prandtl numbers reveals a clear power-law dependence, with slopes steeper than those typically found in purely hydrodynamic convection \citep[e.g.,][] {Ahlers_et_al_2009, Chilla_et_al_2012, He_et_al_2012, Urban_et_al_2012, Iyer_et_al_2020, Pandey_Sreenivasan_EPL_2021}. 
This difference is probably due to the kind of turbulence obtained here, where the correlation between temperature and velocity field fluctuations increases with increasing Re, Rm, and Ra parameters more than in the other studies. The differences in power-law indices between the non-magnetized fluid and an MHD flow are not surprising if we consider how the magnetic field can significantly change the behavior of many hydrodynamic instabilities \citep{Chandrasekhar_book}. In particular, in our systems the magnetic field turns out to increase the correlation between temperature and velocity fluctuations, effectively enhancing the efficiency of convective heat transport in strongly turbulent regimes.

In this framework, our results naturally connect with the classical studies on the asymptotic laws of convective heat transfer in non-magnetic fluids \citep[e.g.,][]{Malkus_54a, Malkus_54b, Spiegel_1962, Kraichnan_1962, Niemela_et_al_2006}. 
However, the presence of magnetic fields introduces additional nonlinear couplings that alter the classical scaling exponents and energy pathways, extending the theoretical framework of turbulent convection into the magnetized domain. 
In particular, we find that in the asymptotic limit, where $Nu$ becomes large, the system also tends to exhibit more frequent magnetic polarity reversals, suggesting a coupling between magnetic variability and the vigor of convective transport.

Our model, although it is a low dimensional model and lacks geometric complexity, describes the main dynamical and nonlinear features of an MHD fluid. It conserves the main physical invariants in the ideal case (i.e., the non-dissipative and unforcing case) and retains the essential nonlinear interactions among velocity, temperature, and magnetic field fluctuations without requiring excessive computational effort. This makes it a valuable theoretical tool for exploring parameter spaces inaccessible to global 3D simulations, guiding the interpretation of more realistic simulations and observations, and for identifying potential scaling regimes relevant to stellar and planetary interiors.

\begin{acknowledgements}
This work is supported by the MELODY research project, funded by the Italian Ministry of Universities and Research (MUR) under the PNRR Young Researchers program 2022 (MELODY SoE project, grant agreement No SOE\_0000119, CUP E53C22002450006).
\end{acknowledgements}

\bibliographystyle{aa}
\bibliography{biblio.bib}
%
\end{document}